## Title: Global maps of the magnetic field in the solar corona

**Authors:** Zihao Yang[1], Christian Bethge[2,3], Hui Tian[1,4*], Steven Tomczyk[3*], Richard Morton[5], Giulio Del Zanna[6], Scott W. McIntosh[3], Bidya Binay Karak[7], Sarah Gibson[3], Tanmoy Samanta[8,9], Jiansen He[1], Yajie Chen[1,10], Linghua Wang[1]

**Affiliations:**

[1]School of Earth and Space Sciences, Peking University, Beijing 100871, People's Republic of China.

[2]Universities Space Research Association, Huntsville, AL 35805, USA.

[3]High Altitude Observatory, National Center for Atmospheric Research, Boulder, CO 80307, USA.

[4]Key Laboratory of Solar Activity, National Astronomical Observatories, Chinese Academy of Sciences, Beijing 100012, People's Republic of China.

[5]Department of Mathematics, Physics & Electrical Engineering, Northumbria University, Newcastle Upon Tyne NE1 8ST, UK.

[6]*Department of Applied Mathematics and Theoretical Physics*, Centre for Mathematical Sciences, University of Cambridge, Cambridge CB3 0WA, UK.

[7]Department of Physics, Indian Institute of Technology (Banaras Hindu University), Varanasi 221005, India.

[8]Department of Physics and Astronomy, George Mason University, Fairfax, VA 22030, USA.

[9]Johns Hopkins University Applied Physics Laboratory, Laurel, MD 20723, USA.

[10]Max Planck Institute for Solar System Research, 37077 Göttingen, Germany.

*Corresponding authors. Email: huitian@pku.edu.cn; tomczyk@ucar.edu

**Abstract:** Understanding many physical processes in the solar atmosphere requires determination of the magnetic field in each atmospheric layer. However, direct measurements of the magnetic field in the Sun's corona are difficult to obtain. Using observations with the Coronal Multi-channel

Polarimeter, we have determined the spatial distribution of the plasma density in the corona, and the phase speed of the prevailing transverse magnetohydrodynamic waves within the plasma. We combine these measurements to map the plane-of-sky component of the global coronal magnetic field. The derived field strengths in the corona from 1.05 to 1.35 solar radii are mostly 1-4 Gauss. These results demonstrate the capability of imaging spectroscopy in coronal magnetic field diagnostics.

**Main Text:**

The solar atmosphere is shaped by its magnetic field. Due to magnetic coupling between the various atmospheric layers, understanding many physical processes in the solar atmosphere requires information on the magnetic field of the whole atmosphere. However, only limited measurements are available for the magnetic field in the upper solar atmosphere, especially in the outermost atmospheric layer (corona) (*1*).

Information on the magnetic field at the solar surface is usually obtained through the Zeeman effect, the splitting of spectral lines in the presence of a magnetic field. However, it is difficult to use this method to measure the coronal magnetic field, mainly due to the negligible line splitting induced by the much weaker magnetic field in the corona. A few attempts have been made to measure the coronal magnetic field through the Zeeman effect, but only in small regions of strong field (*2, 3*). Spectro-polarimetric measurements can also determine the local coronal magnetic field in some cool loop-like structures or prominences (e.g. *4, 5*). Coronal magnetic field strengths can be inferred from observations of waves and oscillations, though previous studies only provided an estimate of the average field strengths in individual oscillating structures (e.g., *6-9*). Observations of shocks driven by solar eruptions can also be used to infer coronal magnetic field strengths along the shock paths (e.g., *10, 11*), but such shocks are only occasionally observed. Radio observations have also been used to estimate the coronal magnetic field, but only in localized regions (e.g., *12, 13*); this method requires accurate identification of the radio emission mechanisms, which are not always clear. Due to the observational difficulties with each of these methods, no routine measurements of the global coronal magnetic field are available.

We used the Coronal Multi-channel Polarimeter (CoMP) (*14*) to observe the corona outside the whole disc of the Sun on 2016 October 14. The CoMP data included spectral profiles of the Fe XIII lines at 1074.7 and 1079.8 nm in the corona from 1.05 to 1.35 solar radii ($R_s$) (*15*). We fitted each line profile with a Gaussian function, then obtained the line intensity and Doppler velocity at each pixel within the CoMP field of view (FOV) (*16*). Figure 1B-C show the intensity images of these two lines averaged over the period of 19:24 Universal Time (UT) to 20:17 UT. For comparison, Fig. 1A shows a simultaneous coronal image in the Fe XII 19.3 nm channel of the Atmospheric Imaging Assembly (AIA) (*17*) on the Solar Dynamics Observatory (SDO) spacecraft. The intensity ratio of the two Fe XIII lines (Fig. 1D) is sensitive to the electron density, allowing us to derive the global coronal electron density map (*15*) (Fig. 1E). The measured electron number density ($N_e$) is mostly in the range of $10^{7.5}$ to $10^{8.5}$ cm$^{-3}$. The associated uncertainties, which arise from both the statistical measurement uncertainties and the systematic uncertainties in the atomic physics parameters used to calculate the relationship between electron density and line ratio (*15*), are mostly 10-25% (Fig. 1F). Assuming a standard coronal elemental abundance and electrical neutrality, the corresponding total mass density ($\rho$) was calculated as $\rho = 1.2 N_e m_p$, where $m_p$ is the mass of a proton (*18*).

Previous CoMP observations have found propagating periodic disturbances in the Doppler velocity of Fe XIII 1074.7 nm, indicating the ubiquitous presence of transverse magnetohydrodynamic (MHD) waves in the corona (e.g., *19-22*). A wave-tracking technique has previously been developed to track the propagation of the Doppler velocity perturbation and calculate the phase speed of the transverse wave along its propagation path (*20*). Similarly pervasive velocity fluctuations also appear in our dataset (Movie S1). We applied a modified version of the wave-tracking technique (*15*) to the Doppler velocity image sequence of Fe XIII 1074.7 nm during the time period of 20:39 UT to 21:26 UT, and calculated the wave phase speed and measurement uncertainty (*15*) at each pixel within the FOV (Fig. 2). The phase speed mostly falls in the range of 300 to 700 km s$^{-1}$, and the associated uncertainty is generally smaller than 40 km s$^{-1}$.

We identify the observed transverse MHD waves as kink waves, which have an Alfvénic nature (e.g., *8, 23-26*). The phase speed (kink speed), $c_k$, can be expressed as (*27*):

$$c_k^2 = \frac{B_i^2 + B_o^2}{\mu_0(\rho_i + \rho_o)} \qquad (1)$$

where $\mu_0$ is the magnetic permeability of a vacuum, $B$ is the magnetic field strength, $\rho$ is the mass density, and the subscripts i and o indicate physical parameters inside and outside the wave-guiding magnetic field structures (flux tubes), respectively. In the coronal plasma environment, the pressure balance across flux tubes is dominated by the magnetic pressure, so $B_i \sim B_o$ (e.g., *20, 21, 26*). Because individual flux tubes are likely unresolved at the spatial resolution of CoMP (~7000 km), we take the density averaged inside and outside flux tubes ($\langle \rho \rangle$) within each spatial pixel, and estimate the magnetic field strength via (*21, 26, 28*):

$$c_k = \frac{B}{\sqrt{\mu_0 \langle \rho \rangle}} \qquad (2)$$

Our measurements are based on spectral profiles that result from an integration of the spectral line emissivity (the released energy per unit time per unit volume during an electron transition from a higher energy level to a lower one, increasing with density) along the line of sight (LOS). The derived density, phase speed and magnetic field strength are therefore all weighted by the emissivity along the LOS. Because the density generally decreases with distance from the solar limb, the LOS weighting favors magnetic structures in the vicinity of the plane of sky (POS), i.e., the plane passing through the center of the Sun and perpendicular to the LOS. We expect the phase speed measured from the data to correspond to the kink speed projected onto the POS. If we further approximate the average density in the vicinity of the POS with the derived density, we can obtain the POS component of the coronal magnetic field strength ($B_{\text{POS}}$) using Eq. 2. Forward simulations of propagating Alfvénic waves have shown that this is an appropriate approximation (*28*).

Our derived global coronal magnetic field map and its uncertainty are shown in Figure 3. Comparing Fig. 3A to the intensity images (Fig. 1, A to C) shows that the magnetic field is higher in regions with stronger coronal emission. Typical values of $B_{\text{POS}}$ in the FOV are $1 - 4$ Gauss (G), similar to the magnetic field strengths in smaller coronal regions inferred using other methods (*3, 10, 11*). The uncertainties on $B_{\text{POS}}$, which we calculated by propagating the uncertainties in the measured density and phase speed (*15*), are shown in Fig. 3B; they are generally smaller than 15%. There could be an additional uncertainty due to our use of the POS emissivity, instead of the LOS-integrated emissivity, in the calculation of the theoretical relationship between line ratio and electron density. As the electron density distribution along the LOS is unknown, we estimated the impact of this assumption using a model of homogeneous density distribution with spherical symmetry (*15*). The density estimated with our line ratio method was lower than the local density in the POS (from the density model) by ~30%. Following Eq. 2, this corresponds to a possible additional uncertainty of ~12% on the measured $B_{\text{POS}}$.

In the absence of routine measurements of the coronal magnetic field, the potential field source surface (PFSS) model (*15, 29*) is often adopted to extrapolate the observed magnetic field on the solar surface to the corona. For comparison with our method, we also used the PFSS model to reconstruct the three-dimensional coronal magnetic field structures from observations of the Helioseismic and Magnetic Imager (HMI) (*30*) on SDO (Fig. 4A) and obtain a map of $B_{\text{POS}}$ from the model (*15*) (Fig. 4B). A comparison between the maps of $B_{\text{POS}}$ extrapolated using the PFSS model and estimated from our data shows similar distributions of coronal magnetic field on the global scale, but differences at scales smaller than ~200 arcsec. At many locations the radial variation of $B_{\text{POS}}$ has a discrepancy between CoMP measurements and PFSS results (Fig. 4, C to F). Some of these differences may arise because the magnitude of $B_{\text{POS}}$ from the PFSS model is plotted for a POS slice. While the $B_{\text{POS}}$ derived from our CoMP data represents a measurement weighted by the emissivity along the LOS, and is the POS component of the magnetic field strength averaged inside and outside flux tubes. Nevertheless, it remains interesting to compare the two $B_{\text{POS}}$ maps since the LOS weighting favors magnetic structures in the vicinity of the POS. Differences between the two $B_{\text{POS}}$ maps could also be related to the assumptions used in the PFSS model (*15*).

Our method for measuring the coronal magnetic field requires a continuous observation of 1-2 hours under good conditions, including ~1 hour to observe the transverse waves and additional time for density diagnostics. This implicitly assumes that the coronal structures do not evolve during the observing period. We expect this to be valid in the absence of eruptive events. The technique cannot be applied to regions affected by solar eruptions, where signatures of transverse waves are masked by the rapidly changing magnetic field environment.

Subject to these assumptions and limitations, our results demonstrate that imaging spectroscopy can be used to determine the coronal magnetic field. In principle, this technique could be applied to continuous observations from CoMP-like instruments to produce routine global coronal magnetic field maps.

**Acknowledgments:** This material is based upon work supported by the National Center for Atmospheric Research, which is a major facility sponsored by the National Science Foundation under Cooperative Agreement No. 1852977. The AIA and HMI are instruments on SDO, a mission of NASA's Living With a Star Program. We thank the SDO team for providing the AIA and HMI data, Michael Galloy for running the CoMP data processing pipeline, and Marc DeRosa for helpful

discussion about PFSS. **Funding:** Supported by NSFC grants 11825301, 11790304(11790300), 41421003, 41774183, 41861134033 and 41874200, Strategic Priority Research Program of CAS (grant XDA17040507), the Max Planck Partner Group program, STFC (UK) via the consolidated grants to the atomic astrophysics group at DAMTP at the University of Cambridge (ST/P000665/1 and ST/T000481/1), and the Department of Science and Technology (SERB/DST) through the Ramanujan Fellowship (project No. SB/S2/RJN-017/2018). **Author contributions:** H.T. led the project. S.T. developed the CoMP instrument, designed the observing sequences and processed the raw data. Z.Y. analyzed the data, generated the figures and movies under H.T.'s guidance. H.T. and Z.Y. wrote the manuscript. C.B improved the CoMP data processing pipeline and led the first stage of data analysis. R.M. contributed to the magnetic field determination. G.D.Z. contributed to the density estimation. S.T. and S.W.M. developed the wave tracking method, with contributions from R.M. and B.B.K. S.G. developed the software for the calculation of POS magnetic field from the PFSS model. Y.C. assisted with the PFSS calculation. T.S., J.H. and L.W. advised on the data analysis and contributed to the interpretation of the observations. All authors have discussed the results and commented on the manuscript. **Competing interests:** There are no competing interests. **Data and materials availability:** The CoMP data can be obtained at http://download.hao.ucar.edu/d5/mlso/pub/comp-continuum-correction/; we used the files that begin with 20161014. The AIA data are available at the Joint Science Operations Center (JSOC) http://jsoc.stanford.edu/AIA/AIA_lev1.html; we used the AIA 193 Å image taken at 19:25:55 UT on 2016 October 14. The HMI synoptic photospheric magnetogram and results of our PFSS model are available in Hierarchical Data Format at http://www.lmsal.com/solarsoft/archive/ssw/pfss_links_v2/Bfield_20161014_180328.h5. Source codes for the wave tracking and density diagnostic are at https://doi.org/10.5281/zenodo.3884044.

**Supplementary Materials:**

Materials and Methods

Supplementary Text

Figs. S1 to S4

Movie S1

References (*31-53*)

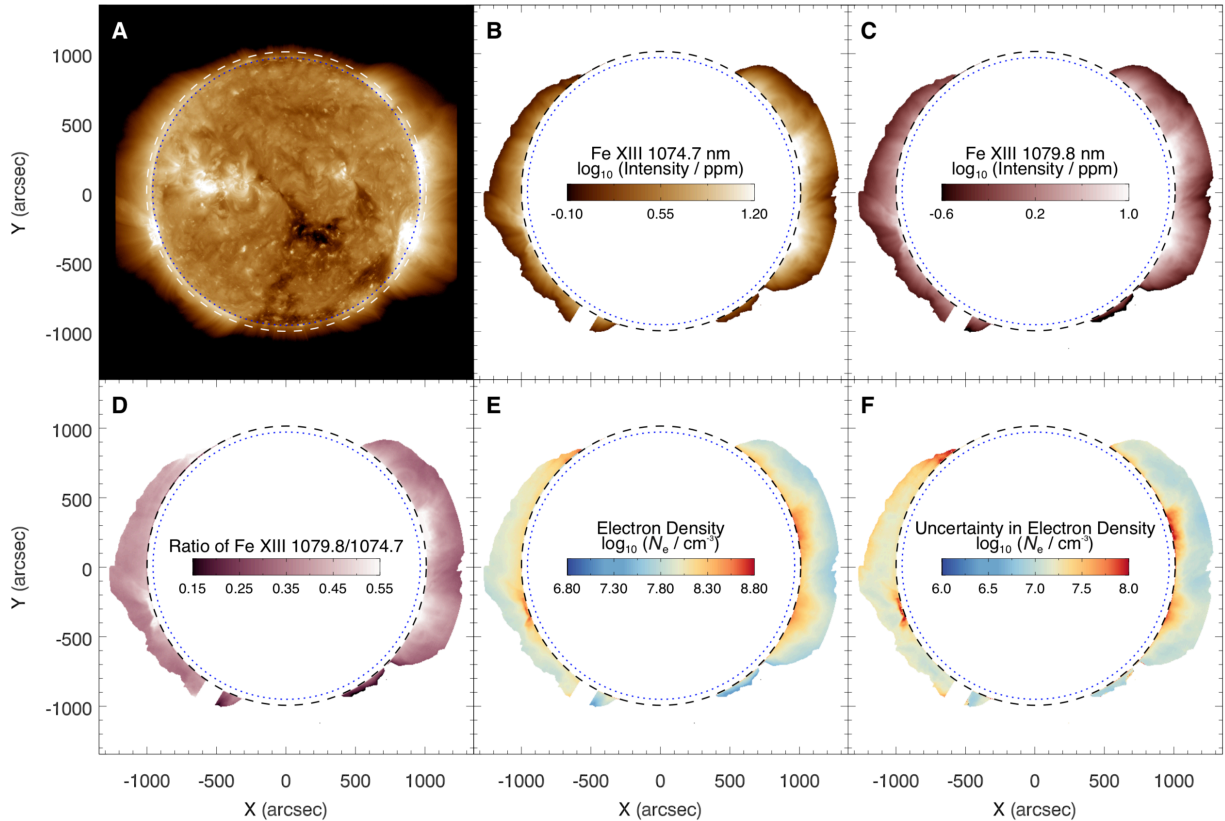

**Fig. 1. Images of the solar corona and density diagnostic results. (A)** AIA 19.3 nm intensity image taken at 19:25:55 UT on 2016 October 14. **(B** and **C)** CoMP Fe XIII 1074.7 nm and 1079.8 nm peak intensity images averaged over the time period of 19:24 UT to 20:17 UT on 2016 October 14, in parts per million (ppm) of the solar disk intensity and plotted on a logarithmic color scale. **(D)** Map of the 1079.8 nm/1074.7 nm intensity ratio. **(E** and **F)** Maps of the derived electron density and associated uncertainty. In all panels, the dotted circle marks the edge of the solar disc (solar limb) and the dashed circle indicates the inner boundary of the CoMP FOV. The X and Y coordinates represent spatial positions in the east-west and south-north directions, respectively.

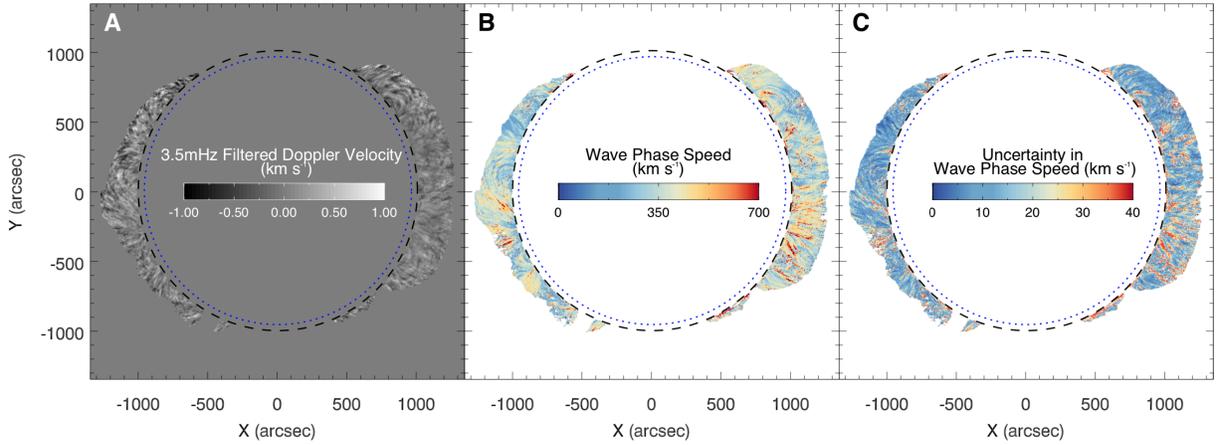

**Fig. 2. Doppler velocity and wave-tracking results. (A)** Map of the Doppler velocity of the Fe XIII 1074.7 nm line at 20:39:09 UT. A 3.5 mHz Gaussian filter has been applied to the Doppler shift image sequence (*15*). Movie S1 shows an animated version of this panel. **(B** and **C)** Maps of the derived wave phase speed and associated uncertainty. The circles are as in Fig. 1.

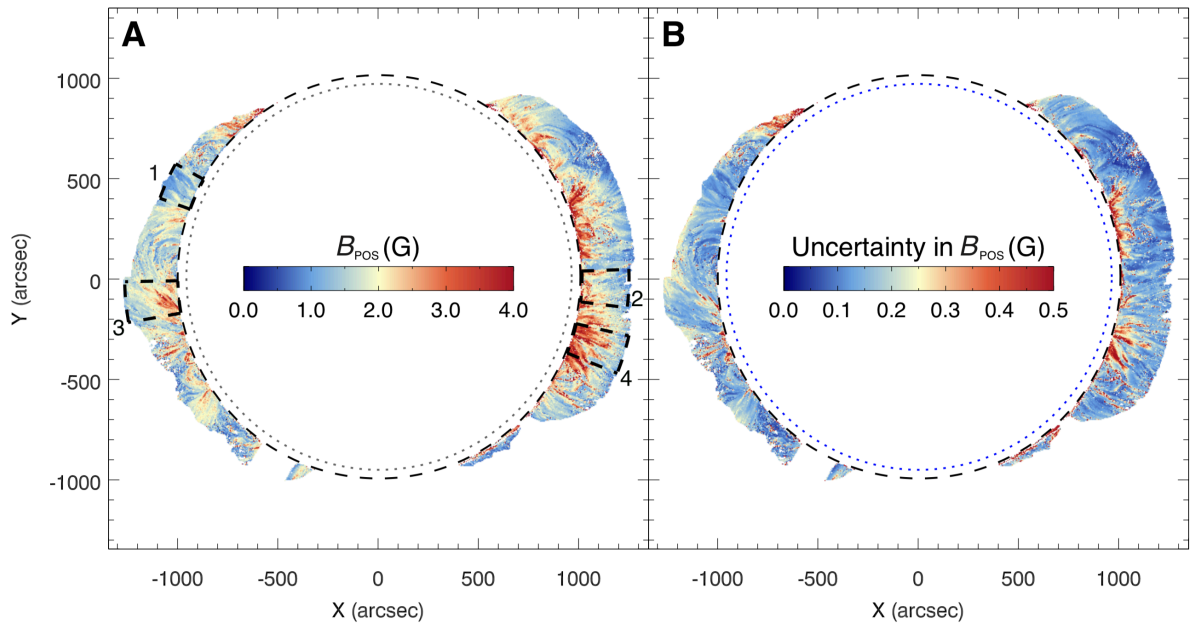

**Fig. 3. Maps of the coronal magnetic field derived from the observations. (A)** Map of the plane-of-sky component of the coronal magnetic field strength ($B_{\text{POS}}$). The four numbered black annulus sectors indicate the regions used for Fig. 4C-F. **(B)** Map of the associated uncertainty. The circles are as in Fig. 1.

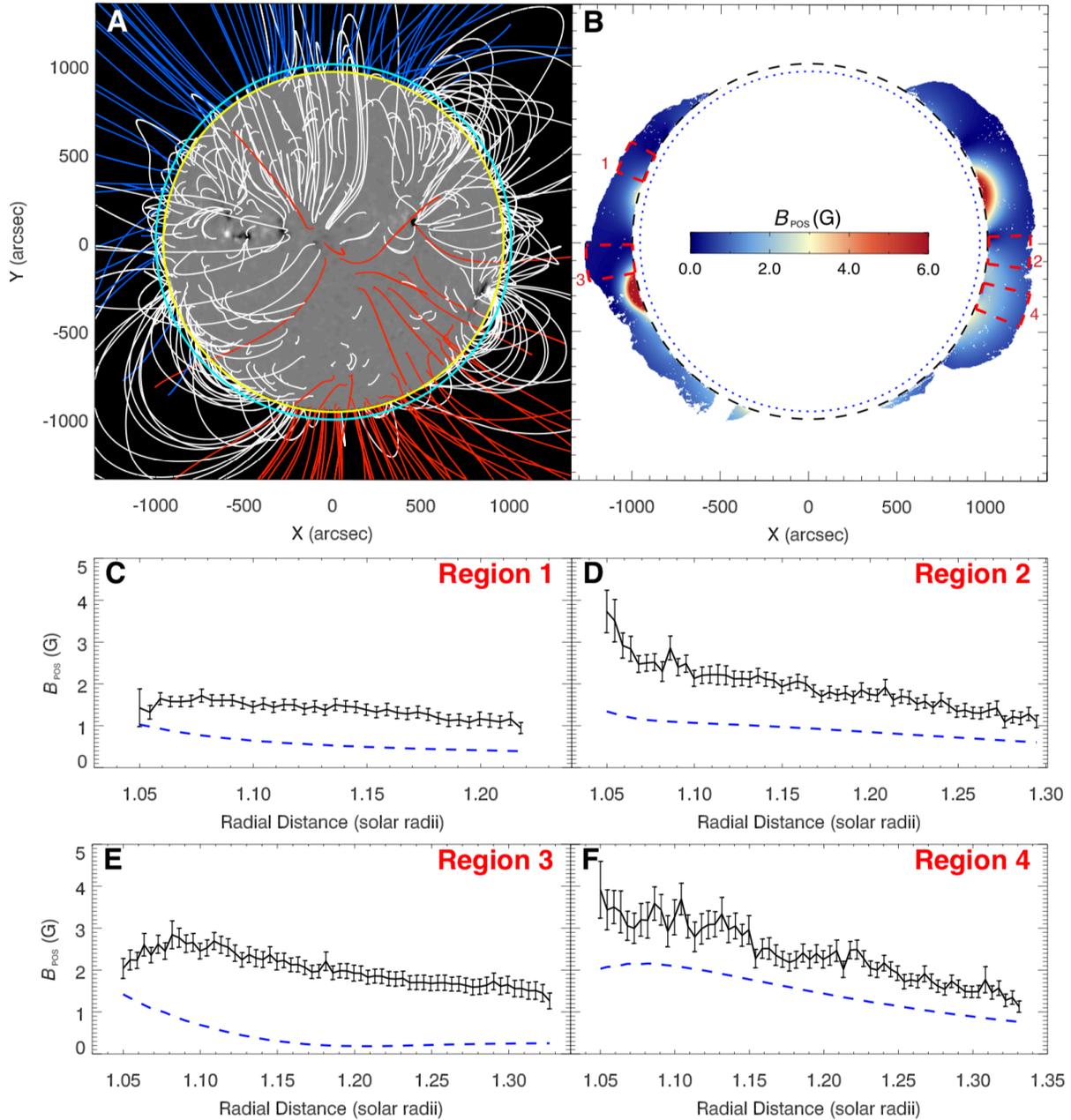

**Fig. 4. Comparison between the coronal magnetic field derived from the observations and extrapolated using the PFSS model. (A)** PFSS model field lines overlain on a photospheric synoptic magnetogram (*15*) reconstructed using SDO/HMI observations. The magnetogram and model field lines sampled at 18:03:28 UT have been rotated and are shown from the Earth's viewpoint at 20:39:09 UT. The yellow and cyan circles mark the solar limb and the inner boundary

of the CoMP FOV, respectively. The white lines are closed field lines, and the red and blue lines represent open field lines with opposite polarities. **(B)** Same as Fig. 3A, but showing the map of $B_{\mathrm{POS}}$ generated from the PFSS model. **(C-F)** Average magnetic field strengths as a function of radial distance from the solar center for the four sectors marked in Fig. 3A and Fig. 4B. The black solid lines with error bars are $B_{\mathrm{POS}}$ derived from the observations and associated uncertainties, and the blue dashed lines show $B_{\mathrm{POS}}$ calculated from the PFSS model.

# Supplementary Materials for

## Global maps of the magnetic field in the solar corona

Zihao Yang, Christian Bethge, Hui Tian[*], Steven Tomczyk[*], Richard Morton, Giulio Del Zanna, Scott W. McIntosh, Bidya Binay Karak, Sarah Gibson, Tanmoy Samanta, Jiansen He,      Yajie Chen, Linghua Wang

*Corresponding authors. Email: huitian@pku.edu.cn; tomczyk@ucar.edu

**This PDF file includes:**

Materials and Methods

Supplementary Text

Figs. S1 to S4

Caption for Movie S1

**Other Supplementary Materials for this manuscript:**

Movie S1

# 1. Materials and Methods

## 1.1 Observations

### 1.1.1 Coronal Multi-channel Polarimeter (CoMP)

We used the data obtained with CoMP on 2016 October 14. CoMP is a coronagraph with the capability of spectropolarimetry at infrared wavelengths (*14*). It can obtain images of the full set of Stokes parameters (*I, Q, U, V*) at several wavelength positions across the profiles of the Fe XIII 1074.7 nm and 1079.8 nm spectral lines from ~1.05 $R_s$ to ~1.35 $R_s$, with a spatial pixel size of ~4.35" through a tunable filter with a full-width-at-half-maximum (FWHM) of 0.14 nm. By taking data with a calibrated diffuser in front of the coronagraph every day, the intensities are normalized and recorded in the unit of ppm (parts per million, one millionth of the solar disk intensity). In our observation, we used the Stokes-*I* (intensity) profiles of the two Fe XIII lines that are formed at typical coronal temperatures (~1.6 million kelvin) under the assumption of ionization equilibrium, as well as the Stokes *Q* and *U* data of the stronger Fe XIII 1074.7 nm line.

The Fe XIII 1074.7 nm images at three wavelength positions (1074.50 nm, 1074.62 nm, 1074.74 nm), taken from 20:39 UT to 21:26 UT, were used for wave tracking. For each wavelength position, there are 94 frames with a time cadence of ~30 s. By assuming a Gaussian shape of a coronal line profile, we could determine the peak intensity, line width and line centroid. Since the line profile at each spatial pixel only has three data points, the three parameters can be calculated from analytical solutions (analytical Gaussian fitting) (*16*). By comparing the centroid with the rest wavelength of the line, we obtained the Doppler velocity at each spatial pixel within the FOV. We used published methods (*16*) for the east-west trend correction and wavelength calibration. The resulting Doppler velocity image sequence was then used for wave tracking and phase speed derivation.

For the density diagnostic, nearly simultaneous Fe XIII 1074.7 nm and 1079.8 nm data during the period of 19:24 UT to 20:17 UT were used. We chose this period because it was close to the wave tracking period of 20:39 UT to 21:26 UT, when no 1079.8 nm data was taken. To increase the signal to noise ratio (S/N), we averaged 49 frames of Stokes-*I* images for the 1074.7 nm line and 10 frames for the 1079.8 nm line at each wavelength position (Fe XIII 1074.7 nm line: 1074.50 nm, 1074.62 nm, 1074.74 nm; Fe XIII 1079.8 nm line: 1079.66 nm, 1079.78 nm, 1079.90 nm). We then obtained the line peak intensity from each averaged line profile by fitting a Gaussian function. A slight shift of the FOV (1 pixel in the east-west direction and 2 pixels in the north-south direction) was found between the wave tracking period and the density diagnostic period. This offset was applied to the data taken during 19:24 UT to 20:17 UT.

To derive the POS direction of magnetic field (azimuth) from linear polarization signals, we used the Fe XIII 1074.7 nm linear polarization (Stokes *Q* and *U*) data sampled at the central wavelength position (1074.62 nm) from 20:39 UT to 21:26 UT. To increase the S/N, 94 images were averaged for both Stokes *Q* and *U*.

To reduce the effect of low S/N on our results, we only used the pixels where the peak intensity of Fe XIII 1074.7 nm is higher than 1.0 ppm. In addition, for the maps of phase speed and magnetic field strength, we only show pixels where the phase speed is in the range of 0 to 2000 km s$^{-1}$.

### 1.1.2   Solar Dynamics Observatory (SDO)

Observations from the AIA and HMI instruments on the SDO spacecraft were also used in our analysis. The AIA 19.3 nm image shown in Fig. 1A was taken at 19:25:55 UT with a spatial pixel size of ~0.6".

The PFSS model uses synoptic photospheric magnetograms constructed from HMI measurements as the lower boundary condition. A synoptic magnetogram represents global photospheric magnetic field (360° in longitude) mapped in Carrrington coordinates, a coordinate system expressed in longitude and latitude of solar surface, with its prime meridian coinciding with the central meridian of the Sun observed from Earth on Nov. 9, 1853 (*31*). For regions within 60° of the solar disk center in the front-side (the Earth-facing side), the HMI magnetogram from the specified observation time is assimilated (by direct insertion) to the global photospheric magnetogram (*32*). For regions outside the assimilated area, magnetic flux information from several days or weeks earlier is moved horizontally based on large-scale flows such as differential rotation (at different latitudes the Sun rotates at different rates). Hence, the synoptic magnetogram shows the global magnetic field during a full rotation of the Sun, consisting of magnetic field data observed at different times. The near-limb and back-side (the side of the Sun that is opposite to the observer on Earth) data were taken from front-side observations several days or weeks earlier. The PFSS model field is sampled at a cadence of 6 hours. We used the model field sampled at 18:03:28 UT, which was then rotated and shown from the Earth's viewpoint at 20:39:09 UT in Fig. 4A.

## 1.2   Database and toolset

### 1.2.1   CHIANTI atomic database

We utilized the CHIANTI database to derive the coronal electron density. CHIANTI (*33*) is an atomic database for calculation of spectra from astrophysical plasmas. We used version 9.0 of CHIANTI (*34*) to perform the density diagnostic.

### 1.2.2 FORWARD software package

FORWARD is an IDL software package for coronal magnetometry (*35*). This toolset allows us to synthesize coronal observables from input models consisting of the coronal magnetic field structures, which could be compared with specific observations. We chose the reconstructed three-dimensional magnetic field calculated from the PFSS model as input, and used FORWARD to synthesize the POS component of the magnetic field magnitude ($B_{POS}$). The synthesized $B_{POS}$ was then compared to $B_{POS}$ obtained from our CoMP observation.

## 1.3   Methods

### 1.3.1   Density diagnostic method

The density diagnostic method is based on the theoretical relationship between electron density and intensity ratio of the Fe XIII 1074.7 nm and 1079.8 nm lines. When synthesizing intensities of these coronal infrared lines, both collisional excitation (transition from a low to a high energy level through electron collisions) and photo-excitation (transition from a low to a high energy level through photon absorption) should be considered (*36, 37*). We calculated the dependence of line ratio on electron density at each height from 1.05 $R_s$ to 1.35 $R_s$, with a step of 0.01 $R_s$. These theoretical curves were used to derive the electron density from the observed line ratio at each pixel in the FOV.

### 1.3.2   Wave tracking method

The wave tracking method has been described in several previous publications (*20, 21, 38*). Here we briefly summarize this method, and refer to the aforementioned literature for details.

We first aligned different frames of the Doppler velocity image sequence through cross correlation, then interpolated the image sequence in time to achieve a regular cadence of 30 s. Because these transverse waves have a dominant period of around 5 minutes (~3.5 mHz) (e.g., *20, 21*), we applied a fast Fourier transform (FFT) to the time series at each spatial pixel and filtered it using a Gaussian window centered at 3.5 mHz with a FWHM of 1.5 mHz. This process removed some noise and led to a velocity time series cleaner than the unfiltered data. For each pixel in the FOV, we defined a box containing $41 \times 41$ pixels surrounding it, then calculated the coherence (the frequency space equivalent of cross-correlation) between the time series at this pixel and its surrounding pixels. The propagation direction of the transverse wave was determined to be aligned with the elongated region of high coherence. Through this method we obtained a map of wave propagation direction. To reduce the effect of noisy measurements, a median filter over 3×3 pixels was applied to this map.

For each pixel, we defined a wave propagation track with a length of 31 pixels using the derived map of wave propagation direction (the wave track length was automatically shortened for pixels close to the boundaries of the FOV). Then a space-time diagram of the Doppler velocity (temporal evolution of velocity along the wave track) was constructed. A two-dimensional FFT was applied to each space-time diagram, resulting in a $k$-$\omega$ diagram (two-dimensional Fourier power spectrum; $k$: wavenumber; $\omega$: frequency). Further space-time diagrams corresponding to inward and outward propagating waves were obtained by computing an inverse FFT of the positive- and negative-frequency parts of the $k$-$\omega$ diagram, respectively (e.g., *20, 39-41*). From the space-time diagram of the outward propagating wave, we cross-correlated the time series at the center of the track with those at other pixels along the track. The wave phase speed at this pixel was derived by fitting a linear model to the relative position along the track as a function of time lag.

### 1.3.3   Estimation of uncertainties

Equation 2 shows that the uncertainty of the derived magnetic field strength is related to the uncertainties of the measured phase speed and derived density.

The uncertainty on the wave phase speed was calculated from the wave tracking procedure. It is dominated by the uncertainty in the fitted parameter (i.e., the gradient) of the linear least-square fitting (*38, 42*).

The uncertainty on the density arises from a combination of the statistical measurement uncertainties of line intensities and the systematic uncertainty of the theoretical relationship between electron density and line intensity ratio (density-line ratio relationship).

The uncertainty on the peak intensity is propagated from the measurement uncertainties of the spectral intensities at the three wavelength positions. The CoMP data noise consists of photon noise $\sigma_\mathrm{p}$, background noise $\sigma_\mathrm{bg}$ (from background subtraction), readout noise $\sigma_\mathrm{r}$, seeing noise $\sigma_\mathrm{see}$ (seeing-induced image motion noise), dark current and flat field noise (*14, 43*). Each image was the result of an average over $m$ exposures (*14*) ($m$ was read from the image file header). The uncertainties caused by the dark current and flat field noise are negligible (*43*), so were not considered in the calculation. The data noise of the spectral intensity (in unit of photons) at wavelength $i$, $\sigma_{I_i}$, can be expressed as

$$\sigma_{I_i}^2 = \frac{\sigma_p^2 + 2\sigma_r^2 + \sigma_{bg}^2 + \sigma_{see}^2}{m} \qquad \text{(S1)}$$

$\sigma_p$, $\sigma_r$ and $\sigma_{bg}$ were determined from expressions in previous publications (*14, 43*). The seeing noise is $\sigma_{see} = \frac{dI}{dz} * \sigma_z$, where $z$, $\frac{dI}{dz}$ and $\sigma_z$ are the spatial distance, the intensity gradient and the uncertainty induced by residual motions due to seeing, respectively (*43*). We took a value of 0.1 for $\sigma_z$ and varied it by one order of magnitude, and found that the corresponding change in $\sigma_{I_i}$ is negligible. The uncertainty on the spectral intensity expressed in ppm is $\frac{\sigma_{I_i}}{k}$ ppm, with $k = 875$ photons ppm$^{-1}$ being a conversion factor (*14*). As mentioned above, we have averaged $n$ frames ($n$=49 for 1074.7 nm and $n$=10 for 1079.8 nm) to increase the S/N, which reduces the measurement uncertainty on the spectral intensity by a factor of $\sqrt{n}$, i.e., the new measurement uncertainty $\delta_{I_i} = \frac{\sigma_{I_i}}{k\sqrt{n}}$. The analytical Gaussian fitting leads to the following expression of peak intensity $I$ (*16*),

$$I = I_2 e^{(v^2/w^2)} \qquad \text{(S2)}$$

where $I_2$ is the intensity at the center of the three wavelength positions, $v$ and $w$ are the Doppler shift and line width, respectively. The uncertainty on the fitted peak intensity ($\delta_I$) is

$$\delta_I^2 = \left(\frac{\partial I}{\partial I_2}\delta_{I_2}\right)^2 + \left(\frac{\partial I}{\partial v}\delta_v\right)^2 + \left(\frac{\partial I}{\partial w}\delta_w\right)^2 \qquad \text{(S3)}$$

where $\delta_{I_2}$, $\delta_v$ and $\delta_w$ are the measurement uncertainty of $I_2$, uncertainties on the derived Doppler velocity and line width, respectively. Similarly, based on the analytical solutions of $v$ and $w$ (*16*), $\delta_v$ and $\delta_w$ were obtained through propagation of the measurement uncertainties of intensities at the three wavelength positions (e.g., *43*).

Another source of uncertainty on the derived density is the uncertainty of the atomic physics parameters involved in the calculation of the theoretical density-line ratio relationship. Following a previous study (*44*), we modified the collisional excitation rates and Einstein coefficients (transition probabilities of spontaneous radiative decays) for the two Fe XIII forbidden transitions and reasonably set uncertainties of 10% for each rate. As the formation of these forbidden lines is affected by cascading effects from higher levels, we estimated the uncertainties by generating 100 density-line ratio curves. The uncertainty of photo-excitation is too small to affect the density uncertainty, since once adopting a local density model, the contribution of photo-excitation only depends on the local height to the solar surface and the photo-exciting radiation from solar disk, which is well measured and stable at infrared wavelengths. Thus, the uncertainty of photo-excitation was neglected in the calculation. In fig. S1 we show the 100 curves when photo-excitation is not considered. For a given density value, the standard deviation ($\Delta R$) of the 100 line ratio values was then taken as the uncertainty ($\pm\Delta R$) of the theoretical density-line ratio curve at each height.

The total uncertainty of the derived electron density was then calculated from a combination of the measurement uncertainties of the two lines and the uncertainty of the theoretical density-line ratio relationship. As depicted in fig. S2, for a certain line ratio $R_0$, there is a corresponding electron density $N_e$. The uncertainty on $R_0$, $\delta R_0$, can be obtained through propagation of the uncertainties of the two Fe XIII line intensities. For each line ratio value within the uncertainty range, the uncertainty of density-line ratio relationship also introduces a variation of its corresponding electron density; hence from fig. S2 we can see that a range of electron density, from a minimum value ($N_{e,min}$) to a maximum one ($N_{e,max}$), corresponds to a certain observed line ratio. The larger value of $|N_e - N_{e,min}|$ and $|N_{e,max} - N_e|$ was taken as the uncertainty of the derived electron density.

## 2. Supplementary Text

## 2.1 Impact of line-of-sight integration on the density diagnostic

We used the IDL routine *dens_plotter.pro*, available in the CHIANTI software package (*33, 34*), for the density diagnostic. The theoretical relationship between line ratio and electron density calculated using *dens_plotter.pro* only considers the line emissivity in the POS. However, the line intensities we observed are the line-of-sight (LOS) integration of the emissivities. We have considered the LOS integration using the following method.

i) We began by assuming the structure of the corona, which is directly related to the distribution of electron density. Following several previous investigations (*36, 45*), we assumed a spherical symmetry for coronal structures. This approach treats the corona as multiple layers of spherical shells. At each radial distance (or shell) above the solar surface the electron density is constant, as given in a one-dimensional model of density as a function of radial distance.

ii) We used a quiet-sun (QS, a region avoiding sunspots and the surrounding areas of strong magnetic field) density model (*46*) as the initial input. By considering both photo-excitation and collisional excitation, the line emissivities were then calculated and integrated along the LOS in a coronal sector by assuming a spherical symmetrical geometry (*45*).

iii) From the LOS-integrated emissivities of the two Fe XIII lines, we obtained the intensity ratio of the two lines as a function of height above the solar limb (fig. S3A), which was then compared with the height variation of the observed line ratio in a chosen QS region (fig. S3C). By adjusting the input density model and ensuring a smooth monotonic decrease of the density with distance, we aimed to match the two curves.

We then compared the adjusted density model and the density derived from the POS line emissivities, and found that the two differ by ~30% at different heights (fig. S3B). The spherical symmetry is a valid assumption only in QS regions. In active regions (ARs) where the corona is highly structured, it is not appropriate to assume a simple three-dimensional geometry for the density distribution. Considering the general agreement in the QS mentioned above, the complexity of ARs, and the fact that a large amount of line emission should come from the vicinity of the POS, we decided to use the density derived from the POS line emissivity in this work. On the basis of the comparison with a QS model with homogeneous distribution with spherical symmetry, we caution that there might be an additional uncertainty of ~30% in the inferred density, due to the unknown distribution of electron density along the LOS. Equation 2 shows that this density uncertainty could lead to an additional uncertainty of ~12% for the derived magnetic field strength.

## 2.2   Potential field source surface model (PFSS)

We compared the measured $B_{\text{POS}}$ with the PFSS model (*29*). Under the potential field assumption, the solar atmosphere above photosphere is current-free ($\nabla \times \boldsymbol{B} = 0$). Therefore, the coronal magnetic field can be expressed using a scalar potential $\Phi$: $\boldsymbol{B} = -\nabla\Phi$ and $\nabla^2\Phi = 0$. The idea of PFSS is to obtain the scalar potential $\Phi$ in spherical coordinate system with known boundary conditions. The lower boundary condition is the photospheric synoptic magnetogram. While the upper boundary is the spherical surface at 2.5 $R_{\text{s}}$ from the center of the Sun, where the coronal magnetic field lines are purely radial. We utilized the PFSS software package (*32*) available in SolarSoft (SSW) to generate the three-dimensional magnetic field.

Figure 4A shows the extrapolated coronal magnetic field lines from the PFSS model. On the global scale, the extrapolated field lines match the observed coronal structures in the AIA 19.3 nm and CoMP intensity images. But there are several regions where they do not match, which could be related to the potential field assumption and the use of a synoptic magnetogram. The PFSS model is based on the assumption that no electric currents exist in the atmosphere, which is likely not the case in some regions of the corona. In addition, the synoptic magnetogram was constructed using HMI measurements at different times during a solar rotation (~27 days), e.g., the photospheric magnetic field data at the east limb was observed several weeks earlier. The magnetic field might have evolved during this period. As a result, the coronal magnetic field extrapolated from this synoptic magnetogram may deviate from the actual coronal magnetic field structures.

A comparison of $B_{\text{POS}}$ generated from FORWARD using the PFSS model (Fig. 4B) with our CoMP observation also reveals a similarity for coronal magnetic field strength on the global scale. However, in many local regions the two show large differences. As discussed in the main text, the different definitions of $B_{\text{POS}}$ are likely responsible for some of the differences. Further differences may be related to the potential field assumption and the use of a synoptic magnetogram in the PFSS model. The discrepancy between our measured magnetic field and the PFSS model result appears to be largest at the east limb (Fig.4, C and E), which might be related to the fact that the east-limb data of the synoptic photospheric magnetogram is several weeks out of date. The non-potentiality of the coronal field could also result in some differences between the two $B_{\text{POS}}$ maps, because the CoMP-derived measure has sensitivity to coronal currents while the PFSS extrapolation does not. Several past studies have compared the magnetic field determined from kink oscillations in coronal loops with known input magnetic field or PFSS field (e.g., *47-50*). These studies generally found a ~20-50% difference in the field strength, consistent with the discrepancy between the two $B_{\text{POS}}$ curves in our Fig. 4, D and F.

## 2.3 Comparison between the wave propagation direction and magnetic azimuth

The magnetic azimuth in the POS ($\phi$) was derived from the averaged Stokes $Q$ and $U$ signals using the relationship

$$\phi = \frac{1}{2}\tan^{-1}\left(\frac{U}{Q}\right). \tag{S4}$$

A scatterplot comparing the azimuths derived from linear polarization signals and the computed wave propagation directions is shown in fig. S4. The noise in the computed magnetic azimuth is inversely proportional to the fraction of light that is linearly polarized (*51*). Thus, we only show the data points where the linear polarization degree is larger than 0.06 to reduce the effect caused by unreliable azimuth measurements. The scatterplot shows a correlation between the azimuth and wave propagation direction, supporting our interpretation that the wave propagation directions computed from our wave-tracking method indicate the magnetic field directions in the POS (*19, 20*).

Figure S4 also indicates the regions subject to the 90° Van Vleck ambiguity, meaning that the azimuth is perpendicular to the direction of the POS component of the magnetic field when the angle between the magnetic field and the solar radial direction is larger than 54.74° (*52, 53*). Only a small fraction of data points are close to these two lines, suggesting that the magnetic field geometry in our observation does not favor the presence of the Van Vleck ambiguity.

## 2.4 Comparison with other methods of coronal magnetic field determination

Our measured values of $B_{\text{POS}}$ in the height range of 1.05 $R_{\text{s}}$ to 1.35 $R_{\text{s}}$ from the solar center are mostly in the range of $1-4$ G (Fig. 3). Previous studies have attempted to infer the magnetic field strengths in some localized coronal regions. For instance, coronal wave observations have revealed a magnetic field strength of $1-9$ G in a similar height range in a loop system (*26*). Through observations of shocks driven by solar eruptions, it was found that the magnetic field is about $1.7-2.1$ G between the heights of 1.1 $R_{\text{s}}$ and 1.2 $R_{\text{s}}$ (*11*), and $1.3-1.5$ G between 1.3 $R_{\text{s}}$ and 1.5 $R_{\text{s}}$ (*10*). From spectropolarimetric observations of an active region, a magnetic field strength of about 4 G was found at the height of ~1.1 $R_{\text{s}}$ (*3*). The magnitude of our derived magnetic field appears to be consistent with these previous results of magnetic field diagnostics in the inner corona.

# 3. Supplementary Figures

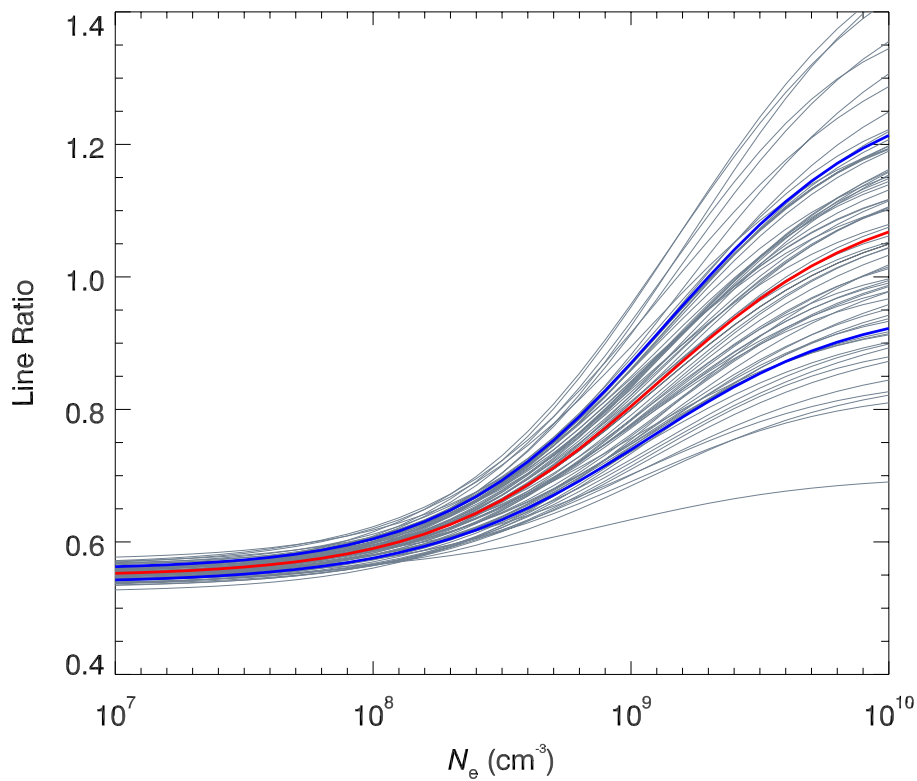

**Fig. S1. Uncertainty of the theoretical density-line ratio relationship.** By setting uncertainties of 10% for the collisional excitation rates and Einstein coefficients associated with the two forbidden transitions, 100 density-line ratio curves were generated (grey curves). Photo-excitation is not considered in this calculation. The red curve shows the mean of the line ratio values at each density. One standard deviation from the mean, as indicated by the two blue curves, has been used for calculation of the uncertainty associated with the theoretical density-line ratio relationship at each height.

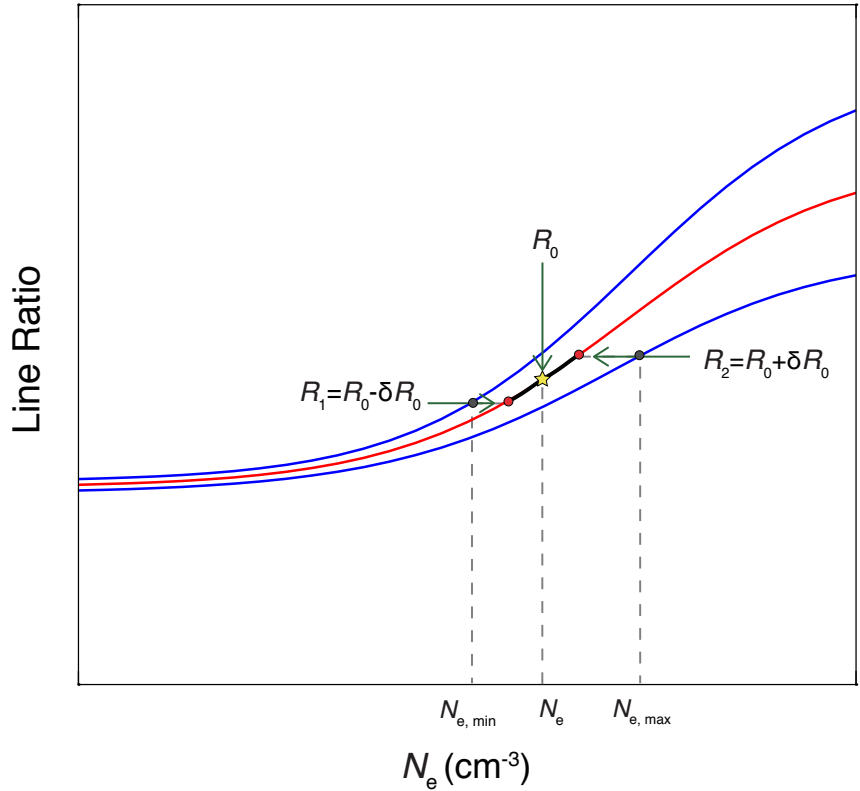

**Fig. S2. A schematic diagram showing estimation of the electron density uncertainty.** The red curve and two blue curves represent the theoretical density-line ratio relationship for a certain height and its uncertainty range (the standard deviation shown in fig. S1). The yellow star represents an observed line ratio $R_0$. The black curve indicates the possible range of actual line ratio value (from $R_1$ to $R_2$) by considering uncertainties of the observed line intensities. The uncertainty of the density-line ratio relationship introduces an uncertainty on the density for any ratio value in this range. As a result, the possible range of electron density that corresponds to $R_0$ is from $N_{e,min}$ to $N_{e,max}$.

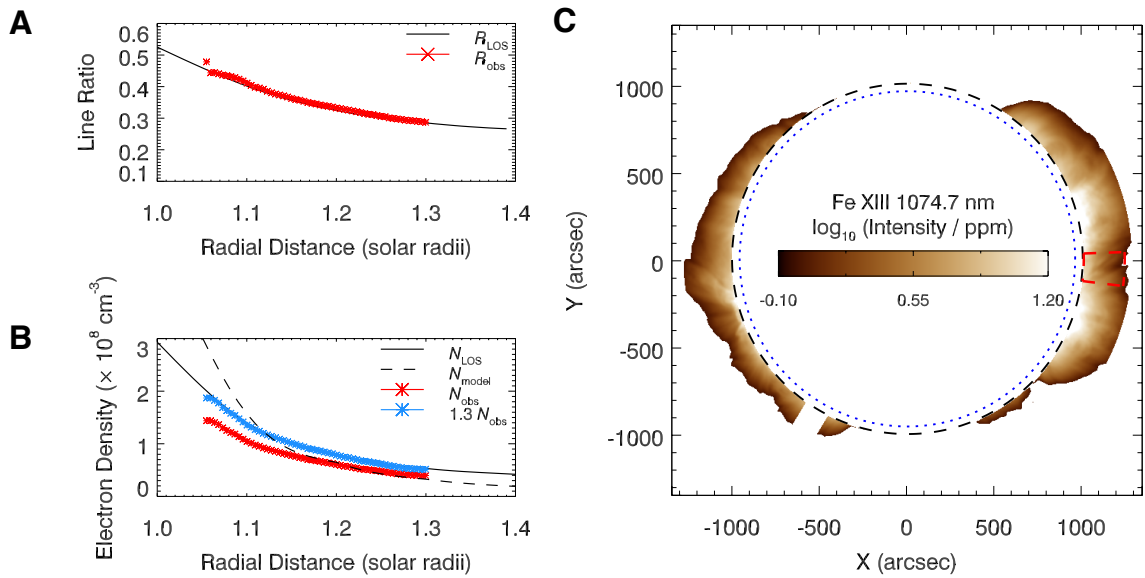

**Fig. S3. Impact of LOS-integration on the density diagnostic. (A)** Height variations of the line ratio derived from LOS-integrated emissivities (solid black line) and from CoMP observation (red stars, from a small QS region marked in panel C). The line ratio derived from LOS integration is matched with the observed value by adjusting the electron density model. **(B)** Comparison between the height variation of the observed electron density (red stars) and the adjusted density model (solid black line). The blue stars represent density values 30% larger than the observed. The dashed line is the input electron density model. **(C)** Same as Fig. 1B. The red sector marked in the west limb is the QS region used as an observational reference in panels A and B.

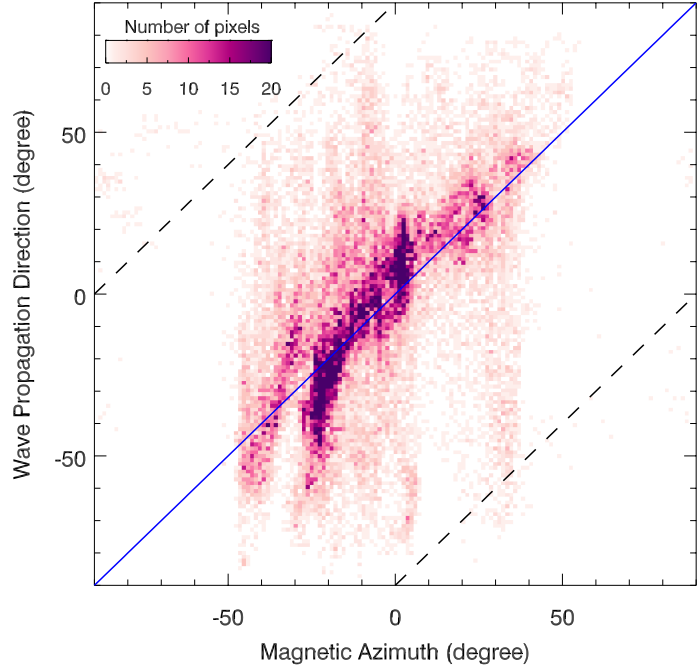

**Fig. S4. Scatterplot showing the relationship between the magnetic azimuth and wave propagation direction.** The number of pixels is represented by the color map. The blue solid line (correlation line) represents the correspondence between azimuth and wave propagation direction. The black dashed lines indicate the regions subject to the 90° Van Vleck ambiguity.

## 4. Caption for Movie S1

The Doppler velocity movie filtered using a Gaussian window centered at 3.5 mHz with a FWHM of 1.5 mHz. The dotted circle marks the solar limb, and the dashed circle indicates the inner boundary of the CoMP FOV. This is an animated version of Fig. 2A.